\documentclass[showpacs,preprintnumbers,amsmath,amssymb,floatfix]{revtex4-2}

\usepackage{graphicx}
\usepackage{dcolumn}
\usepackage{bm}
\usepackage{amssymb}
\usepackage{epsfig}
\usepackage{color}

\newcommand{\ba}{\begin{eqnarray}}
\newcommand{\ea}{\end{eqnarray}}
\newcommand{\be}{\begin{equation}}
\newcommand{\ee}{\end{equation}}
\newcommand{\bdisplay}{\begin{displaymath}}
\newcommand{\edisplay}{\end{displaymath}}

\newcommand{\eq}[1]{Eq.\,(\ref{#1})}
\newcommand{\fig}[1]{Fig.\,\ref{#1}}

\newcommand{\pbar}{\bar{p}}

\makeatletter
\def\eqnarray{\stepcounter{equation}\let\@currentlabel=\theequation
\global\@eqnswtrue
\tabskip\@centering\let\\=\@eqncr
$$\halign to \displaywidth\bgroup\hfil\global\@eqcnt\z@
  $\displaystyle\tabskip\z@{##}$&\global\@eqcnt\@ne
  \hfil$\displaystyle{{}##{}}$\hfil
  &\global\@eqcnt\tw@ $\displaystyle{##}$\hfil
  \tabskip\@centering&\llap{##}\tabskip\z@\cr}

\def\endeqnarray{\@@eqncr\egroup
      \global\advance\c@equation\m@ne$$\global\@ignoretrue}

\def\@yeqncr{\@ifnextchar [{\@xeqncr}{\@xeqncr[5pt]}}
\makeatother

\begin{document}

\title{Simple calculation of the Coulomb-nuclear corrections in $pp$ and $\pbar p$ scattering}

\author{ Loyal Durand}
\email{ldurand@hep.wisc.edu}
\altaffiliation{Present address: 415 Pearl Court, Aspen, CO 81611}
\affiliation{Department of Physics
University of Wisconsin-Madison, Madison, WI 53706}

\author{Phuoc Ha}
\email{pdha@towson.edu}
\affiliation{Department of Physics, Astronomy, and Geosciences, Towson University, Towson, MD 21252}

\begin{abstract}

We present a very simple method for calculating the mixed Coulomb-nuclear effects in the $pp$ and $\bar{p}p$ scattering amplitudes, and illustrate the method using simple models frequently used to describe their differential cross sections at small momentum transfers. Combined with the pure Coulomb and form-factor contributions to the scattering amplitude which are known analytically from prior work, and the unmixed nuclear or strong-interaction scattering amplitude, the results give a much simpler approach to fitting the measured $pp$ and $\pbar p$ cross sections and extracting information on the real part of the forward scattering amplitudes than methods now in use. 

\end{abstract}

\pacs{}

\maketitle


\section{Introduction \label{sec:intro}} 

The effect of the Coulomb interaction in high-energy proton-proton and antiproton-proton  scattering has been studied by many authors; see \cite{BetheCoulomb,WestYennie,Cahn,Islam,KL-Coulomb,KopeliovichTarasov,Petrov1,DH_CoulNucl,Nekrasov} and the many further references therein. A primary objective has been the use of Coulomb-nuclear interference effects to determine the ratios $\rho$ of the real to imaginary parts of the $pp$ and $\pbar p$ scattering amplitudes in the forward direction. The only other direct information on the real parts of the amplitudes is that obtained at much larger angles very near the observed dips in the differential cross sections. These dips are associated with diffraction zeros in the imaginary parts of the amplitudes where the scattering is dominated by the real part \cite{Ha_Re_f}. 

The most commonly used method for calculating the Coulomb-nuclear effects in data analyses appears to be that of Cahn \cite{Cahn}  as later modified by  Kundr\'{a}t and M. Lokaji\v{c}ek  \cite{KL-Coulomb}. This method is based on the use of the Fourier-Bessel convolution theorem to calculate the corrections that involve Coulomb and nuclear interactions simultaneously, and to include the effects of the nucleon charge form factors. The results, which involve delicate manipulations in their derivation to avoid singularities associated with the infinite range of the Coulomb interaction, involve further complications in the subsequent evaluation of convolutions, and are not transparent. 

We show here that the mixed Coulomb-nuclear correction term can in fact be calculated very simply as the rapidly-convergent double integral 
\be
 \label{double_int1}
  f_{N,c}^{\,\rm Corr}(s,q^2) = \int_0^\infty db b\left[ \left(e^{2i\delta'_c(b,s)+2i\delta_c^{FF}(b,s)}-1\right)\times \int_0^\infty dq' q' f_N(s,q'^2)J_0(q'b)\right]J_0(qb),
\ee
where the inner integration on $q'$ must be performed first, as indicated.  Here $\delta'_c(b,s)$ is the pure Coulomb phase shift or eikonal function after an impact-parameter independent term $i\eta\ln{(2p/q)}$ is extracted, and $\delta_c^{FF}(b,s)$ is the phase shift associated with the effects of the charge form factors of the nucleons.   $f_N(b,s)$ is the nuclear scattering amplitude. This also appears as a separate term in the full amplitude as shown below. The remaining Coulomb and form-factor-associated terms in the amplitude are known analytically \cite{DH_CoulNucl}.

In a typical analysis such as that in \cite{TOTEM2016}, the sum of a model nuclear amplitude $f_N$ and the Coulomb amplitude modified by form factors is used to fit the data with the model $f_N$ also used to calculate the rather complicated mixed Coulomb, nuclear, and form-factor terms following Cahn \cite{Cahn} or Kundr\'{a}t and M. Lokaji\v{c}ek  \cite{KL-Coulomb}. The fit is then adjusted as necessary to account for those corrections. Our approach is similar, but with the important differences that we use analytically known expressions \cite{DH_CoulNucl} for the combined Coulomb and form-factor contribution, write the amplitude in a form that isolates a pure nuclear term, and write the remaining mixed Coulomb-nuclear term as in \eq{double_int1}. As we show by example, that correction is small and easily calculated, significantly simplifying the analysis.


\section{Theoretical background \label{sec:theory}}

In the absence of significant spin effects, generally thought to be very small at high energies, the spin-averaged differential cross section  for proton-proton scattering can be written in terms of a single  spin-independent amplitude
\be
\label{f^tot}
f(s,q^2) = i\int_0^\infty db\,b\left(1-e^{2i\delta_{tot}(b,s)}\right)J_0(qb).
\ee
The total phase shift $\delta_{tot}$ the sum of terms $\delta_c$ for pure Coulomb scattering, $\delta_c^{FF}$ for the effects of the charge form factors of the proton, and $\delta_N$ for the strong-interaction or ``nuclear'' scattering,
\be
\label{delta_total}
\delta_{tot}(b,s) = \delta_c(b,s)+\delta_c^{FF}(b,s) + \delta_N(b,s).
\ee
Here
\be
\label{delta_c} 
\delta_c(b,s) = \eta(\ln{pb}+\gamma) 
\ee
where $\gamma=0.5772\ldots$ is Euler's constant,  $\eta=z_1z_2\alpha =\alpha$ $(-\alpha)$ for $pp$ $(\pbar p)$  scattering, and \cite{corr1} 
\be
\label{delta_FF}
\delta_c^{FF}(b,s) = \sum_{m=0}^3\frac{\eta}{2^m\Gamma(m+1)}(\mu b)^mK_m(\mu b)
\ee
for the standard proton charge form factor 
\be
\label{form_factor}
F_Q(q^2) = \frac{\mu^4}{\left(q^2+\mu^2\right)^2}
\ee
with $\mu^2=0.71$ GeV$^2$.

With our normalization, the differential scattering cross section is
\be
\label{dsigma/dq2}
\frac{d\sigma}{dq^2} = \pi\lvert f(s,q^2)\rvert^2,
\ee
where $q^2=-t$ is the square of the invariant momentum transfer and $W=\sqrt{s}$ the total energy in the center-of-mass system.

 As shown in \cite{Ha_Re_f}, an overall momentum-dependent phase factor $(4p^2/q^2)^{i\eta}$ which does not affect the  differential cross section can be extracted from the scattering amplitude, and the remaining amplitude written in the form 
  \be
  \label{f_defined}
  f (s,q^2) = f'_c(s,q^2)+f_c^{FF}(s,q^2)+f_N(s,q^2)+f_{N,c}^{\,\rm Corr}(s,q^2),
  \ee
with $\delta_c\rightarrow \delta'_c$ now given in \eq{delta_total} by 
\be
\label{delta'_c}
\delta'_c(b,s) = \eta(\ln{(qb/2)}+\gamma).
\ee
The Coulomb and form-factor terms combine as shown in \cite{DH_CoulNucl}, Sec.~IIC, to give
  \be
   \label{f_Coul}
   f_c(s,q^2)+f_c^{FF}(s,q^2) = -\frac{2\eta}{q^2}
  \left[1-\left(\frac{q^2}{q^2+\mu^2}\right)^{i\eta}\left(1-\frac{\mu^8}{(q^2+\mu^2)^4}\right)+O(\eta)\right],
  \ee
 with the overall phase $\left(4p^2/q^2\right)^{i\eta}$ which appears  in Eq.~(21) of \cite{DH_CoulNucl} removed.
 
 The purely nuclear amplitude, which is to be determined from fits to scattering data, is
  \be
   \label{f_N}
  f_N(s,q^2) = i\int_0^\infty db b \left(1-e^{2i\delta_N(b,s)}\right)J_0(qb).
  \ee
Finally,
  \be
 \label{corr_int}
  f_{N,c}^{\,\rm Corr}(s,q^2) = \int_0^\infty db b\left(e^{2i\delta'_c(b,s)+2i\delta_c^{FF}(b,s)}-1\right)\times i\left(1-e^{2i\delta_N(b,s)}\right)J_0(qb)
  \ee
  is the mixed Coulomb-nuclear term.
  
  We note that the last factor in \eq{corr_int} is just the integrand for $f_N$ in \eq{f_N}, so may be evaluated as the inverse Fourier-Bessel transform $\tilde{f}_N(b,s) $ of the nuclear amplitude $f_N$ \cite{Bessel_inverse},
  \be
  \label{inv_Bessel}
\tilde{f}_N(b,s) = \int_0^\infty dq q f_N(s,q^2)J_0(qb) =  i\left(1-e^{2i\delta_N(b,s)}\right). 
 \ee
Thus,
  \be
  \label{corr_int2}
    f_{N,c}^{\,\rm Corr}(s,q^2) = \int_0^\infty db b\left(e^{2i\delta'_c(b,s)+2i\delta_c^{FF}(b,s)}-1\right)\tilde{f}_N(b,s).
    \ee

The key observation is that $\tilde{f}_N(b,s)$ can be determined for any successful phenomenological model for $f_N(s,q^2)$ by performing the inverse transform in \eq{inv_Bessel}.  This can be calculated analytically for the exponential-type models in $q^2$ commonly used in fitting the $pp$ and $\bar{p}p$ data at high energies and small momentum transfers, and some other models as well,   giving simple expressions that make the calculation of the Coulomb-nuclear correction straightforward by numerical evaluation of the remaining rapidly-convergent integral.  We will consider some examples in Sec.~III.

The mixed Coulomb-nuclear correction can also be evaluated efficiently numerically for models in which $\tilde{f}_N(b,s)$ cannot be calculated analytically. In that case,
\be
\label{double_int2}
  f_{N,c}^{\,\rm Corr}(s,q^2) = \int_0^\infty db b\left[ \left(e^{2i\delta'_c(b,s)+2i\delta_c^{FF}(b,s)}-1\right)\times \int_0^\infty dq' q' f_N(s,q'^2)J_0(q'b)\right]J_0(qb)
\ee
as stated in the Introduction.
 The key is to first evaluate the integral over $q'$ in \eq{double_int2}. This integral converges rapidly for any reasonable model for $f_N$ that describes the rapid, nearly exponential, fall of the differential cross sections with increasing $q^2$ observed at high energies, and gives a result that  vanishes rapidly for large $b$ as  expected from the long-range behavior of strong interactions. The second integral over $b$ is therefore also expected to converge rapidly. 
 This will be seen explicitly in the examples in Sec.~III.
 
 The order of the integrations is crucial: the Coulomb plus form factor term in parentheses in \eq{double_int2} does not provide convergence at large $b$ if one tries to integrate in the opposite order, and one encounters the singularities that caused trouble in Cahn's approach \cite{Cahn} and its later modifications. The double integral in \eq{double_int2} converges well overall when performed in the order specified, and can easily be evaluated numerically. 
  
 In our approach, as in that of Cahn \cite{Cahn} or Kundr\'{a}t and M. Lokaji\v{c}ek  \cite{KL-Coulomb},   or the various modifications of it, one has to start with a model for $f_N$ fitted to the data, then calculate the Coulomb-nuclear correction, which will be small, and then refit to get the final $f_N$.


\section{Examples \label{sec:examples}}

Consider as an example the simple exponential model
\be
\label{exp_model}
f_N^{\,\rm exp}(s,q^2)=(i+\rho)\sqrt{A/\pi}e^{-\frac{1}{2}Bq^2}
\ee
with $A,\,B$ and $\rho$ functions of $s$ but independent of $b$.  This has been used over a wide range of energies to fit experimental data on the $pp$ and $\pbar p$ differential cross sections at small $q^2$ to determine the forward slope parameters $B=-d(\ln{d\sigma/dq^2})/dq^2$, the total cross sections $\sigma_{\rm tot}= 4\pi\Im f_N(s,0)$, and to determine the ratios $\rho(s)=\Re f_N(s,0)/\Im f_N(s,0)$ of the real to the imaginary parts of the forward amplitudes from Coulomb-nuclear interference effects. See, for example, \cite{Amos_ISR} and \cite{TOTEM2016} for examples at 52.8 GeV and 8 TeV.  For this model
\be
\label{exp_model2}
\tilde{f}_N^{\,\rm exp}(b,s) = (i+\rho)\int_0^\infty dq q \sqrt{\frac{A}{\pi}}e^{-\frac{1}{2}Bq^2}J_0(qb)=(i+\rho)\sqrt{\frac{A}{\pi}}\frac{1}{B}e^{-b^2/2B}.
 \ee
The remaining integral over $b$ in \eq{double_int2} converges exponentially and is easily evaluated numerically  to get the mixed Coulomb-nuclear correction term  $f_{N,c}^{\,\rm Corr}(s,q^2)$. 

 The model can be extended to
\be
\label{fNwithzero}
f_N^{\,\rm exp'}(s,q^2) = (i+\rho)\sqrt{A/\pi}e^{-\frac{1}{2}B'q^2}(1-cq^2)
\ee
 to include the diffraction structure caused by the expected zero in $\Im f_N$. Here $c=1/q_0^2$, $q_0^2$ is the location of the observed diffraction minimum, and  $B'=B-2c$ is fixed to reproduce the observed forward slope $B$.   In this case, 
\be
\label{fNwithzero2}
\tilde{f}_N^{\,\rm exp'}(b,s) = (i+\rho)\sqrt{\frac{A}{\pi}}\frac{1}{B'}\left(1-\frac{2c}{B'}+\frac{cb^2}{B'^2}\right)e^{-b^2/2B'}.
\ee
 We have not found this extension to be necessary in calculating the corrections for $q^2 \ll q_0^2$.

The inner integral in \eq{double_int2} can also be evaluated analytically for some other models, for example,  that of Ferreira, Kohara, and Kodama \cite{Ferreira}, which fits the $pp$ data quite well over a wide range in $q^2$.  However, for most models, the integrals must be evaluated numerically. As shown by the simple exponential model, the inner and outer integrals may be expected to converge very rapidly as functions of $q'$ and $b$ for realistic $f_N$.

\begin{figure}[htbp]
\includegraphics{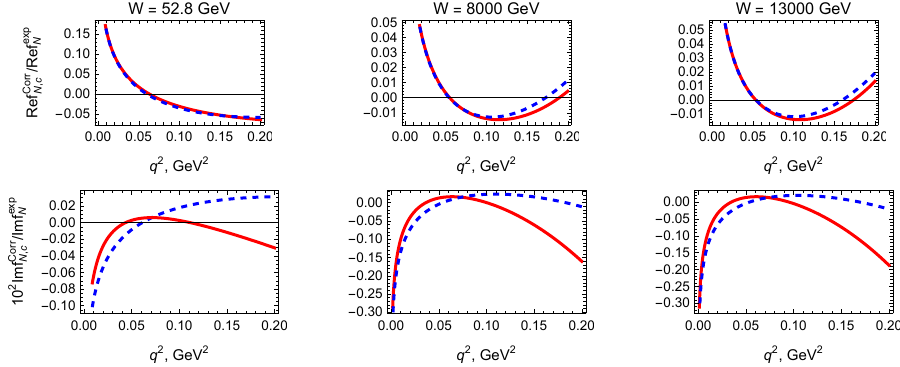}
 \caption{Top row: ratios of the real parts of the mixed Coulomb-nuclear corrections $f_{N,c}^{\,\rm Corr}(s,q^2)$  to the $pp$ elastic scattering amplitude calculated using the eikonal model of Block {\em et al.} \cite{eikonal2015} for the nuclear amplitude $f_N(s,q^2)$ (solid red lines) and the simple exponential model $f_N^{\rm exp}(s,q^2)$ of \eq{exp_model} (dashed blue line), to the real parts of $f_N^{\rm exp}(s,q^2)$. Bottom row: ratios of the imaginary parts of  $f_{N,c}^{\,\rm Corr}(s,q^2)$ calculated using the eikonal model  (solid red line) and the exponential model (dashed blue line) to the imaginary part of the exponential model.  }
 \label{fig1}
\end{figure}

 In \fig{fig1} we show the ratios of the real and imaginary parts of the mixed Coulomb-nuclear corrections $f_{N,c}^{\,\rm Corr}(s,q^2)$ calculated using the eikonal model of Block {\em et al.} \cite{eikonal2015} (red solid curves), and the exponential model of \eq{exp_model} (blue dashed curves), to the  real and imaginary parts of the simple exponential model  $f_N^{\,\rm exp}(s,q^2)$. The full eikonal model satisfies the constraints of unitarity, analyticity, and crossing symmetry, fits the data on $\sigma_{\rm tot}$, $\sigma_{\rm elas}$, $B$, and $\rho$ for $pp$ and $\pbar p$ scattering from 5 GeV to 57 TeV, and gives a good description of the differential scattering cross sections and dip structure even though the data on $d\sigma/dq^2$ other than B were not used in the fit. It is taken here as representing the experimental data. The simple model uses the values of $\sigma_{\rm tot}$, $B$, and $\rho$ obtained in the eikonal fit to obtain $A,\,B$ and $\rho$ in \eq{exp_model} as would be obtained in a fit to the data. 
 
 As seen in the top row in \fig{fig1}, the real parts of the corrections calculated using the simple exponential model and \eq{corr_int2} agree remarkably well at small $q^2$ with those calculated in the eikonal model using the expression in \eq{corr_int} with the eikonal phase shift. This agreement  would be expected. The real part of the correction is associated mainly with the imaginary part of the nuclear scattering amplitude as may be seen by expanding the exponential in the factor in parentheses in \eq{corr_int2} to first order in the small quantity $\eta$. Since $\Im f_N\gg\lvert\Re f_N\rvert$, the very good fit of the exponential model to $d\sigma/dq^2$ over a range of small $q^2$ over which the cross section decreases rapidly as seen, for example, in \cite{Amos_ISR} and \cite{TOTEM2016},  implies a correspondingly good fit  to $\Im f_N$ over a region in which the integral in \eq{corr_int2} converges rapidly.  
 
 The imaginary parts of the mixed Coulomb-nuclear correction found using the exponential model are considerably less accurate, but are  quite small as seen in the bottom row in \fig{fig1}. They arise mainly from the real part of $f_N$, small compared to the imaginary part, and give very small corrections to the latter which are not involved in the Coulomb-nuclear interference, hence, in the determination of $\rho$. 
 
 The relative inaccuracy of these corrections seen in \fig{fig1} results from the poor description of $\Re f_N$ given by the exponential model. As expected from a theorem of Martin \cite{Martin} and seen in the eikonal model, there is a diffraction zero in $\Re f_N$ between $q^2=0$ and the first diffraction zero in $\Im f_N$. This is not evident in the differential cross section because of the small size of $\Re f_N$ relative to $\Im f_N$ for $q^2$ below the dip region, but still leads to a much more rapid decrease of $\Re f_N$ than $\Im f_N$ as $q^2$ increases from 0. The Martin zero is not included in $f_N^{\rm exp}$, \eq{exp_model}, and can only be incorporated in a model like that in \eq{fNwithzero} using information on the position of the zero that is not available from experiment. We have found that simply including a separate exponential term for the real part of the model amplitude with a magnitude $\rho$ relative to the imaginary part and a slope parameter $B_R$ matched to that in the eikonal model eliminates most of the errors in the comparisons in the bottom row in \fig{fig1}.


\begin{figure}[th]
\centering
\includegraphics[width=7.5cm]{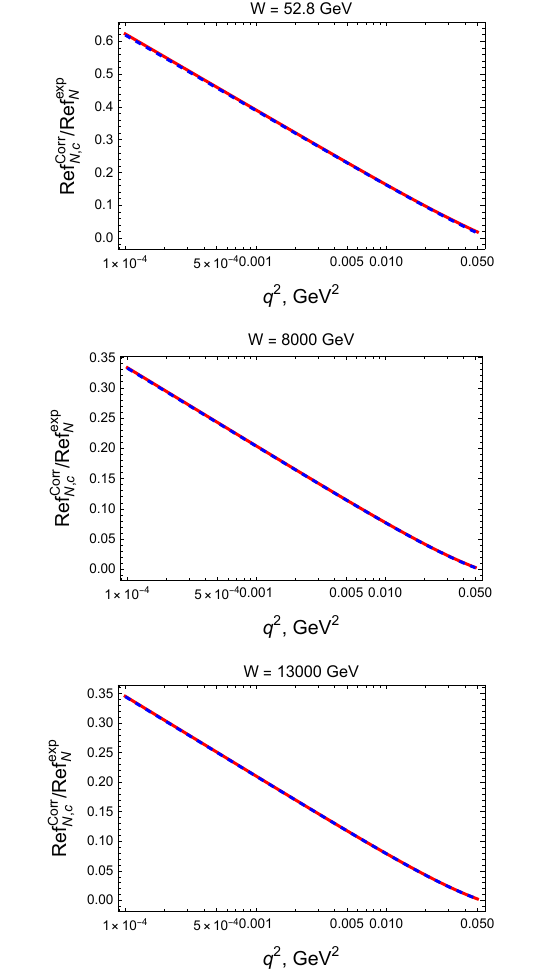}
\includegraphics[width=7.5cm]{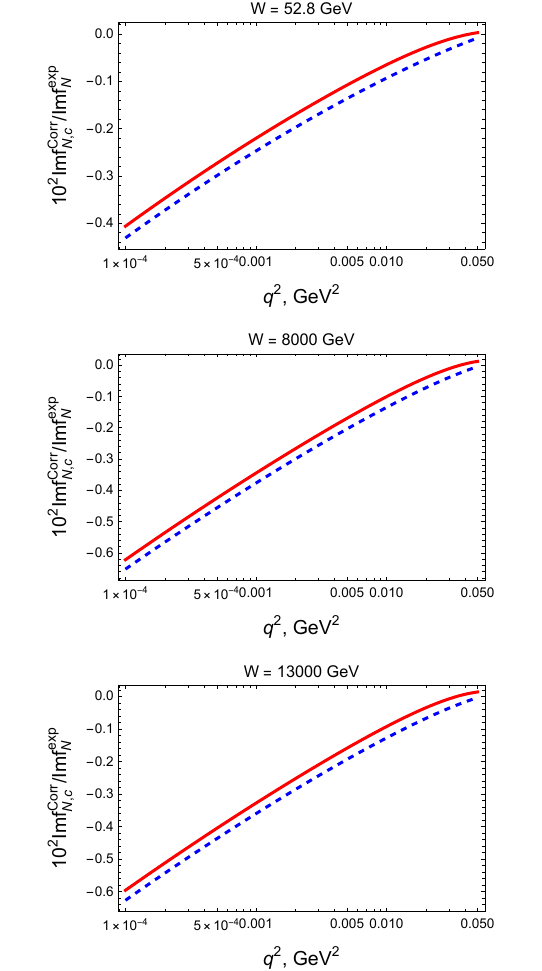}
\caption{Left-hand column: ratios of the real parts of the mixed Coulomb-nuclear corrections $f_{N,c}^{\,\rm Corr}(s,q^2)$  to the $pp$ elastic scattering amplitude calculated using the eikonal model (solid red lines) and the simple exponential model  of \eq{exp_model} (dashed blue line), to the real parts of $f_N^{\rm exp}$ in the region of Coulomb-nuclear interference. Right-hand column: ratios of the imaginary parts of  $f_{N,c}^{\,\rm Corr}(s,q^2)$ calculated using the eikonal model  (solid red line) and the exponential model (dashed blue line) to the imaginary part of the exponential model.  }
\label{fig2}
\end{figure}

In \fig{fig2} we show  ratios of the real and imaginary parts of the mixed Coulomb-nuclear corrections $f_{N,c}^{\,\rm Corr}(s,q^2)$ to the real and imaginary parts of $f_N^{\rm exp}$ for the eikonal and exponential models at very small $q^2$, the region of the observed Coulomb-nuclear interference. The agreement of the results for the real parts is excellent.  The corrections to the real part of the amplitude are substantial and diverge logarithmically at small $q^2$ because of the term $\ln{(qb/2)}$ in $\delta'_c$, \eq{delta'_c}. While small compared to the Coulomb term itself, the corrections in $\Re f_N^{\,\rm Corr}$ cannot be neglected in analyses of Coulomb-nuclear interference.

\begin{figure}[htbp]
\includegraphics{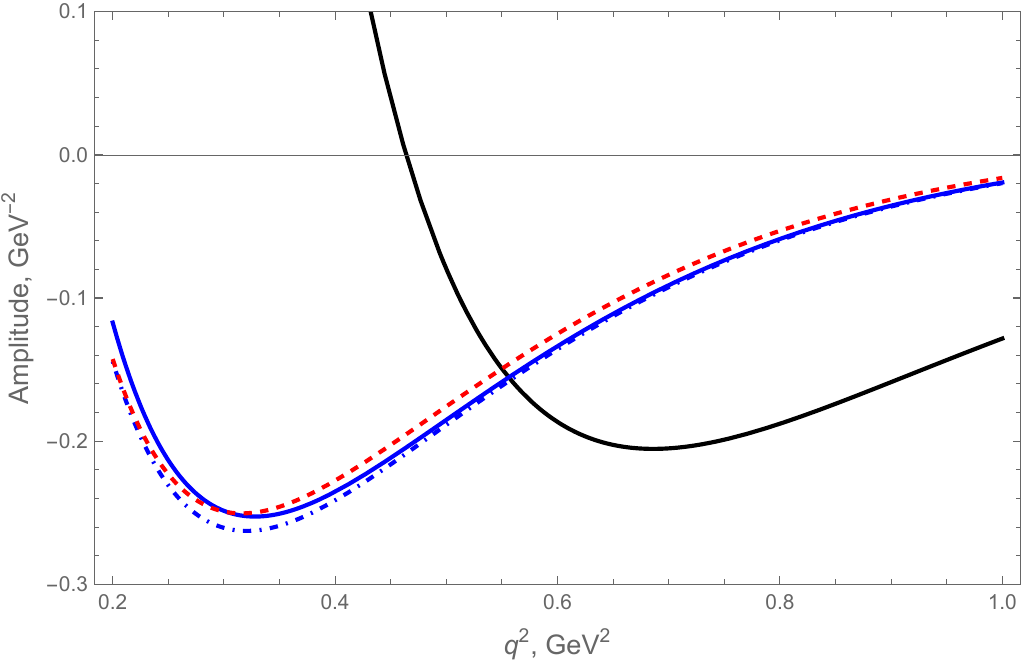}
 \caption{Plots of the nuclear amplitude and corrections in the neighbor hood of the diffraction zero in $\Im f_N$ in the eikonal model of Block {\em et al.} \cite{eikonal2015}  at 8000 GeV. Solid black line: $\Im f_N(s,q^2)$. Solid blue line: $\Re f_N(s,q^2)$. Dash-dot blue line: sum of $\Re f_N$ and the Coulomb plus form-factor term in \eq{f_Coul}. Dashed red line: the real part of the full $pp$ scattering amplitude including the Coulomb and form-factor terms and the mixed Coulomb-nuclear corrections.}
 \label{fig3}
\end{figure}

In \fig{fig3}, we show the real and imaginary parts of $f_N$ in the neighborhood of the diffraction zero in $\Im f_N$ for $pp$ scattering at 8 TeV as calculated  in the eikonal model. The real part of the amplitude is dominant near the zero, and the Coulomb term as modified by the charge form factors of the protons leads to an approximately 20\% difference between the $pp$ and $\pbar p$ cross sections even with no significant crossing-odd hadronic term present in $f_N$ \cite{Ha_Re_f}. The mixed Coulomb-nuclear corrections to $\Re f_N$ are again small but significant. The magnitude of the differential cross section at the minimum provides a measure of $\Re f_N$ at that point, where it is known to be negative since $\rho$ is positive as measured for $q^2\approx 0$ and $\Re f_N$ changes  sign at the Martin zero \cite{Martin}.

\section{Conclusions \label{sec:conclusions}}
 
We conclude that the very simple calculation of the Coiulomb-nuclear corrections to the $pp$ and $\pbar p$ scattering amplitudes  through \eq{corr_int2} or \eq{double_int2} using a reasonable model for $f_N$ is quite accurate at high energies, and represents substantial improvement  in simplicity and clarity over the methods most commonly used at present \cite{Cahn,KL-Coulomb,Petrov1}. The combination of the analytic expression for the pure Coulomb and form-factor terms in \eq{f_Coul}, the model nuclear amplitude $f_N$, and this small correction term gives our expression for the full $pp$ or $\pbar p$ scattering amplitude $f(s,q^2)$, \eq{f_defined}. 

We have illustrated the calculation of the mixed Coulomb-nuclear term for $pp$ scattering using the simple exponential model for the nuclear scattering amplitude in \eq{exp_model} which is frequently used to fit cross sections at  high energies and small $q^2$, and shown that the results agree with those obtained in the comprehensive eikonal model of Block {\em et al.} \cite{eikonal2015} at energies from $W=52.8$ GeV to 13 TeV. A useful parametrization of the  correction term is given in \cite{Ha_Re_f}, Sec.~II. We have not investigated the corrections at low energies, but the method still applies for appropriate models for $f_N$.

%
%


\begin{thebibliography}{99}
\bibitem{corr1} The expression given for $2\delta_c^{FF}$ in Eq.~(17) of \cite{DH_CoulNucl} is missing the factor 2 and has the wrong sign and summation limits on the right-hand side as given there.
\end{thebibliography}

\end{document}